\documentclass[preprint]{aastex}

\newcommand\Msun{\ensuremath{\mathrm{M_\odot}}}
\newcommand\Rsun{\ensuremath{\mathrm{R_\odot}}}
\newcommand\Lsun{\ensuremath{\mathrm{L_\odot}}}
\usepackage{epstopdf}
\usepackage{amsmath}
\usepackage{bbding}

\begin{document}
\title{Keck Adaptive Optics Observations of the Protostellar Disk around Radio Source I in the Orion Kleinmann-Low Nebula}

\author{\sc Breann N. Sitarski\altaffilmark{1}, Mark R. Morris\altaffilmark{1}, Jessica R. Lu\altaffilmark{2}, Gaspard Duch\^{e}ne\altaffilmark{3, 4}, Andrea Stolte\altaffilmark{5}, E. E. Becklin\altaffilmark{1, 6}, Andrea M. Ghez\altaffilmark{1}, Hans Zinnecker\altaffilmark{6, 7, 8}}

\email{bsitarski@astro.ucla.edu}

\altaffiltext{1}{Department of Physics and Astronomy, University of California, Los Angeles, 430 Portola Plaza, Los Angeles, CA 90095-1547, USA}
\altaffiltext{2}{Institute for Astronomy, University of Hawaii, 2680 Woodlawn Drive, Honolulu, HI 96822}
\altaffiltext{3}{Department of Astronomy, University of California, Berkeley, C-203  Hearst Field Annex, Berkeley, CA 94720-3411}
\altaffiltext{4}{UJF - Grenoble 1/ CNRS-INSU, Institut de Plan\'etologie et d'Astrophysique de Grenoble (IPAG) UMR 5274, Grenoble, F-38041, France}
\altaffiltext{5}{Argelander Institut f\"{u}r Astronomie, Universit\"{a}t Bonn, Auf dem H\"{u}gel 71, 53121 Bonn}
\altaffiltext{6}{NASA-Ames Research Center, MS 232-12, Moffett Field, CA 94035, USA}
\altaffiltext{7}{Astrophysikalisches Institut Potsdam, An der Sternwarte 16, 14482 Potsdam, Germany}
\altaffiltext{8}{Deutsches SOFIA Institut, Universit\"{a}t Stuttgart, Pfaffenwaldring 31, 70569 Stuttgart, Germany}

\begin{abstract}

We have made the first detection of a near-infrared counterpart associated with the disk around Radio Source ``I," a massive protostar in the Kleinmann-Low Nebula in Orion using imaging with laser guide star adaptive optics on the Keck II telescope. The infrared emission is evident in images acquired using $L^{\prime}$ (3.8 $\mu$m) and $Ms$ (4.7 $\mu$m) filters and is not detectable at $K^{\prime}$ (2.1 $\mu$m). The observed morphology strongly suggests that we are seeing some combination of scattered and thermal light emanating from the disk. The disk is also manifest in the $L^{\prime}/Ms$ flux ratio image. We interpret the near-infrared emission as the illuminated surface of a nearly edge-on disk, oriented so that only the northern face is visible; the opposite surface remains hidden by the disk. We do not see infrared radiation associated directly with the star proposed to be associated with Source ``I." The data also suggest that there is a cavity above and below the disk that is oriented perpendicular to the disk, and is sculpted by the known, strong outflow from the inner disk of Source I. We compare our data to models of a protostar with a surrounding disk, envelope, and wind-blown cavity in order to elucidate the nature of the disk around Radio Source I.

\end{abstract}


\section{Introduction}
At a distance of $415$ $\pm$ $6$ pc \citep{kim2008, menten2007}, the Kleinmann-Low Nebula (KL) in Orion is one of the nearest sites of massive star formation \citep{menten2007}. The KL nebula hosts at least three, possibly more, massive protostars and the tremendous luminosity of the KL nebula ($L \sim 10^{5}$ $\Lsun$, \citealt{genzel1989, werner1976}) suggests that one or more of these protostars is dramatically influencing its surrounding gas and dust. Owing to the extreme extinction toward this region ($A_{v}>30$ magnitudes; \citealt{lagrange2004}), the star-formation there has primarily been targeted by infrared and radio studies. The massive, luminous sources embedded in dust have elucidated the collapse phase of what will become massive stars. 

The KL region plays host to several heavily-reddened sources of kinematic and radiative energy that may be the powering sources of the region. Among these is the Becklin-Neugebauer object (BN), and the compact thermal radio sources `I' and `n'. The BN object, first recognized as a massive protostar by \citet{becklin1967}, has been classified as a B0 star \citep{scoville1983}. NICMOS polarization studies have revealed that BN heats its surrounding disk environment but is likely not the source of heating for the nebula \citep{simpson2006}. Several other bright, diffuse infrared sources may also contribute to the luminosity, including IRc 4 \citep{debuizer2012}; however, no stellar point sources have been detected. Radio sources `I' and `n' are thought to be associated with the formation of at least two other massive stars \citep{mentenandreid1995}, and there are numerous other sources bright at near- and mid-infrared wavelengths (e.g. \citealt{dougados1993}; \citealt{gezari1998}; \citealt{shuping2004}). In fact, Source I may be the dominant source of energy in the KL region.

The nature of Source I has been under debate in recent years. It is completely obscured at all infrared wavelengths out to at least 37 $\mu$m \citep{debuizer2012}, and is only detected as a color temperature peak in a 7.8 $\mu$m/12.4 $\mu$m color-temperature analysis \citep{okumura2011}. \citet{mentenandreid1995} showed that the radio continuum measured from Source I directly coincides with the centroid of SiO maser emission seen at 43.1 GHz, suggesting that Source I is the source powering the maser emission. This hypothesis was further supported by 87 GHz observations of thermal SiO emission \citep{plambeck2009}, detected as a large bipolar outflow extending from the northeast to the southwest for at least eight arcseconds. The outflow has also been seen in submillimeter line emission (e.g., \citealt{beuther2005}; \citealt{beuther2008}, \citealt{zapata2012}). The 3-mm SiO maser emission occurs at the base of the outflow in an \textbf{${\times}$}-shaped pattern \citep{matthews2010}. There has not been a universal consensus on the orientation of the disk and outflow \citep{bally2011}, but \citet{matthews2010} argue rather compellingly that the  semi-major axis of the disk must be oriented northwest to southeast, and this is currently the predominant paradigm.

The obscuration of Source I may, in large part, be due to obscuration by its own edge-on disk, centered on the radio point source. Both \citet{reid2007} and \citet{goddi2011} have imaged Source I in 7-mm radio continuum, which shows that the radio extension is perpendicular to the outflow revealed by SiO emission. This led \citet{reid2007} and \citet{goddi2011} to suggest that the radio emission arises from the ionized surface of the inner disk. The detection at near- to mid-infrared wavelengths of the disk surrounding the radio point source has been elusive due to the high extinction in the region \citep{greenhill2004a}. Characterizing the protostellar disk surrounding Source I would allow us to paint a more complete picture of massive star formation, since very few disks around young, massive stars have been characterized satisfactorily (e.g., \citealt{zapata2006}, \citealt{okamoto2009}).

The only previous high-angular resolution infrared imaging of the region around Source I -- the IRc 2 region -- was presented by \citet{dougados1993} (see Figure 1 for the proximity of Source I to the IRc 2 region). With an average spatial resolution of $\sim$ 0."5, they were able to resolve the four components of the IRc 2 region (IRc 2A, 2B, 2C, and 2D; see Figure 7 in that paper) using a deconvolution method on data taken at the 4 meter Kitt Peak Mayall telescope.

We present $K^{\prime}$, $L^{\prime}$, and $Ms$ imaging of the BN/KL region using laser guide star adaptive optics on the Keck II 10-m telescope, the deepest and highest-resolution near-infrared imaging of the region to date. This deep imaging allows us to construct intensity and color-temperature maps of the region that show the infrared counterpart to the disk around Radio Source I. We are able to achieve an average angular resolution of 0."09 at $L^{\prime}$. These high angular resolution data allow us to distinctively see the extended emission and to resolve the point sources in the region.

Section 2 of this paper discusses the observations; the data reduction and analysis procedures are presented in Section 3. Section 4 presents our primary finding, and Section 5 provides a discussion of the properties of the disk around Radio Source I. We summarize our conclusions and the implications for future work in Section 6.

\section{Observations}

The Orion BN/KL star-forming region was observed using laser guide star adaptive optics (LGSAO; \citealt{wizinowich2006}, \citealt{vandam2006}) on 2010 October 30 - November 1 at the W. M. Keck II 10-meter telescope with the facility near-infrared camera, NIRC2 (PI: K. Matthews). These images were taken with the narrow-field camera, which has a pixel scale of 10 mas pixel$^{-1}$ and a field of view of $\sim$ 10\arcsec $\times$ 10\arcsec \citep{ghez2008}. Observations were taken through the $K^{\prime}$ ($\lambda_{0}$ = 2.12 $\mu m$, $\Delta\lambda$ = 0.31 $\mu m$), $L^{\prime}$ ($\lambda_{0}$ = 3.78 $\mu m$, $\Delta\lambda$ = 0.70 $\mu m$), and $Ms$ ($\lambda_{0}$ = 4.67 $\mu m$, $\Delta\lambda$ = 0.24 $\mu m$) filters. The images were dithered randomly within a total offset region of 0."7 $\times$ 0."7. Table 1 summarizes the observational details for each data set obtained. The natural seeing was $\sim$ 0.5 arcseconds during both nights. The laser guide star  corrected for most of the atmospheric aberrations, but the low-order tip-tilt terms were corrected using visible light observations of Parenago 1839 ($\alpha_{J2000}$ = 05:35:14.64, $\delta_{J2000}$ = -05:22:33.7; $R$ = 13.8; see Figure 1 for placement in our field of view). Sky frames at $L^{\prime}$ and $Ms$ were taken interspersed with science observations in a dark region $\sim15$ degrees to the East. Sky observations were timed such that the field rotator mirror angle was identical to that of the science exposures. This was necessary to accurately subtract thermal emission from the field rotator mirror, which is in a focusing beam (see \citet{stolte2010} for more details). 

\begin{deluxetable}{ccrrrrrrrrrcrl}
\tabletypesize{\scriptsize}
\tablecaption{Keck II NIRC2 Orion Observations}
\tablewidth{0pt}
\tablehead{\colhead{Date} & \colhead{Filter} &  \colhead{Exposure Time\tablenotemark{a}} & \colhead{Number of} & \colhead{Total Integration}\\
 & & t$_{int}$ $\times$ coadds (sec) & Frames & Time (sec)}
\startdata
2010 Oct 31 &  $K^{\prime}$ & 29.04 & 128 & 3717 \\
2010 Oct 30 &  $L^{\prime}$ & 30.00 & 149 & 4470\\
2010 Nov 1 &  $Ms$ & 27.15 & 31 & 842\\

\enddata
\tablenotetext{a}{The integration time of one frame multiplied by the total number of coadds.}
\end{deluxetable}

\section{Data Analysis}
\subsection{Reduction of Images}
The data were run through the standard reduction pipeline that the UCLA group has developed for reduction of adaptive optics data on the Galactic Center (e.g., \citealt{lu2009}), including dark and flat-field correction, sky subtraction, removal of bad pixels and cosmic rays, correction for differential atmospheric refraction, and application of the NIRC2 distortion map \citep{yelda2010}. For each $L^{\prime}$ and $Ms$ science exposure, a series of sky exposures was subtracted where the sky and science exposures had matching angles on the field rotator mirror, in order to properly subtract any thermal emission from dust on the mirror optics. For the $K^{\prime}$ science exposures, the skies were combined into a single image used for background subtraction. Images were registered and combined using the IRAF/PyRAF modules \textit{xregister} and \textit{drizzle}. The reduced data typically had strehl ratios of $\sim$ 0.25, 0.5, and 0.52 and FWHMs of $\sim$65, 85, and 105 mas at $K^{\prime}$, $L^{\prime}$, and $Ms$, respectively. The individual images were combined into both the final average image as well as three subimages, each comprised of one-third of the data in order to estimate photometric and astrometric uncertainties. 

\subsection{Locating Source I in the Near-Infrared Images}

Radio Source I has never been detected in the near-infrared, so its position in the near-infrared is unknown. However, the nearby source, ``n", is detected at both radio and infrared wavelengths and is used to astrometrically register our infrared images to the radio reference frame. 

To extract precise astrometry for all sources within our infrared images, we use the IDL package \textit{StarFinder} \citep{diolaiti2000}. \textit{StarFinder} iteratively extracts a point spread function (PSF) from a set of user-selected stars and identifies point sources within the field of view. We implement a version of \textit{StarFinder} that has been modified to trim out sources that are likely to be artifacts of imperfect PSF knowledge, as described in \citet{yelda2010}. Our narrow-field images had very few bright point sources, so only four were used as reference PSF sources -- Source n \citep{lonsdale1982}, Parenago 1839  \citep{strand1958}, $[$SCE2006$]$ 14 and 15 \citep{hillenbrand2000} (see Figure 1 for their placement within the field of view). These sources were selected because they were far enough away from the edge of the detector that they did not suffered minimal flux loss, and they were not immersed in extended emission. The resulting PSF was used to identify candidate sources in the image that exceed a PSF correlation threshold of 0.7 and 0.9 in the subimages and main map respectively. Astrometry and photometry for the candidate sources are computed through PSF fitting. Candidate sources are declared stars if they are detected in the fully combined image and in all subimages. We positively identified seven point sources with a typical astrometric error of 1.5 mas and a photometric error of $\sim$ 20\%. 

Precision astrometry from \textit{StarFinder} was necessary to find the sub-pixel position of Source n in our near-infrared images. We used the near-infrared image position coupled with the known radio proper motion and radio position of Source n (\citealt{goddi2011}; see Table 2 below) to make Source n the origin of our coordinate system. The radio positions in Table 2 are from 13 Nov 2000 \citep{gomez2005}; with the proper motions given in Table 2, we translated those positions to the same 2010 epoch as the NIRC2 data to find the location of Source I relative to Source n. Prior to 2006, Source n was a double in the radio (e.g., \citealt{mentenandreid1995, gomez2008}) with the lobes separated by 0.373 $\pm$ 0.013 arcseconds in April 1994 \citep{mentenandreid1995}, but it appeared as a point source in the near-infrared \citep{dougados1993}. Source n became a single radio object by 2006 \citep{gomez2008}, so we do not need to correct for the ambiguities represented by a double source. 

The proper motion of Radio Source I has been well-studied (e.g., \citealt{goddi2011}; \citealt{gomez2008}). In an effort to reduce the error on any astrometric measurement of Source I, we compiled and combined the absolute positions from \citet{goddi2011} and \citet{gomez2008} and applied a linear least-squares fit to the right ascension and declination, weighted by the uncertainties in the positions of the source. This fit yielded proper motions of $\mu_{\alpha}$ = 4.26 $\pm$ 0.63 mas/year and $\mu_{\delta}$ = -5.52 $\pm$ 0.68 mas/year in the absolute coordinate system (Figure 2 and Table 2). These are well within the uncertainties in the proper motions presented by \citet{gomez2008} ($\mu_{\alpha}$ = 4.5 $\pm$ 1.2 mas/year, $\mu_{\delta}$ = -5.7 $\pm$ 1.3 mas/year) and \citet{goddi2011} ($\mu_{\alpha}$ = 6.3 $\pm$ 1.1 mas/year, $\mu_{\delta}$ = -4.2 $\pm$ 1.1 mas/year).

\begin{figure}[!h]
\begin{center}
\includegraphics[width=15cm]{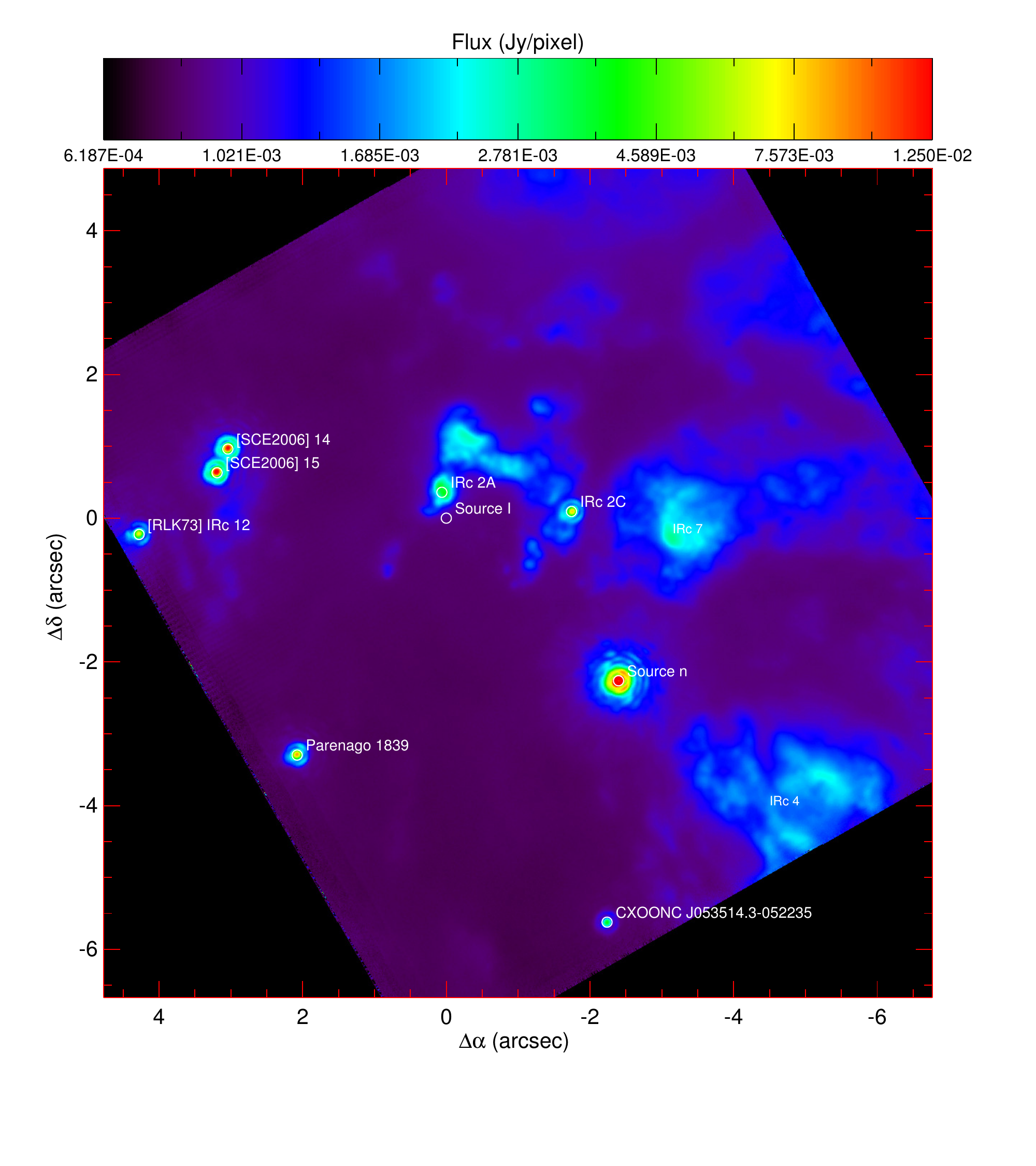}
\caption{NIRC2 $L^{\prime}$ narrow-field image on a logarithmic stretch with overplotted sources. The origin of the coordinate system is at Radio Source I in this image (denoted by the open white circle located at the origin), and the positions of the sources were obtained using \textit{StarFinder}. North is pointing up, and East is to the left.}
\end{center}
\end{figure}

Using Source n as the reference point and the calculated proper motion of Source I, we determined the position of Source I in our NIRC2 images. This position is denoted by the open circle in Figure 1. We verify our coordinate conversion using the separation between the BN object and source n, which are both detected in infrared and radio images. The BN object is typically outside our narrow-view NIRC2 images; however, we obtained a wide-field $K^{\prime}$ image with NIRC2 in 2010 with a 40" $\times$ 40" field of view. After registering our infrared images in the manner described above, the separation between BN and source n is measured as $10\farcs81 \pm 0\farcs04$ at a position angle of 20.4 $\pm$ 0.3$^{\circ}$ measured from Source n to BN. The predicted separation from radio measurements is $10\farcs79 \pm 0\farcs03$ for the same epoch at a position angle of 19.9 $\pm$ 0.2$^{\circ}$ measured from Source n to BN. This agreement shows that our positional estimate for Source I in our infrared observations is accurate to 17 mas.

\begin{figure}[!h]
\hskip-1cm
\includegraphics[width=19.2cm]{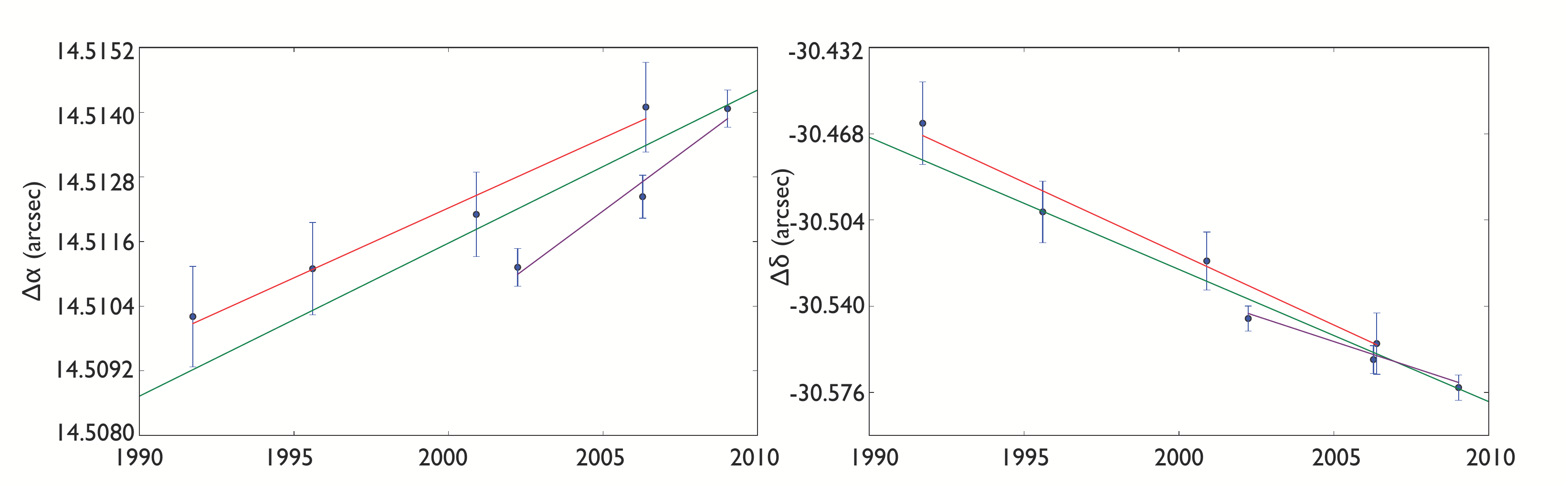}
\caption{\textit{Left}: Absolute right ascension of Radio Source I versus time from \citet{goddi2011} (red line) and \citet{gomez2008} (purple line). The green line represents a least-squares fit to all the data points. \textit{Right}: Absolute declination positions versus time, with the same meaning for the line colors. Taken individually, the two data sets from \citet{goddi2011} and \citet{gomez2008} lead to consistent proper motions within uncertainties. The absolute proper motion derived from these absolute positions is necessary to determine the position of Source I in our 2010 infrared data.}
\end{figure}

\begin{deluxetable}{ccrrrrrrrrrcrl}
\tabletypesize{\scriptsize}
\tablecaption{Radio Astrometry}
\tablewidth{0pt}
\tablehead{\colhead{Source}& \colhead{J2000 Radio Position} & \colhead{Radio Proper Motion} & \colhead{J2000 Position} & \colhead{Reference}\\
 & ($\alpha, \delta$), 13 Nov 2000 & ($\mu_{\alpha}, \mu_{\delta}$) (mas yr$^{-1}$) &  1 Nov 2010\tablenotemark{a} &}
\startdata
Source n  & (05 35 14.3553, -05 22 32.702) & (0.0 $\pm$ 0.9, -13.0 $\pm$ 1.2) & (05 35 14.3553, -05 22 32.832) & \citealt{gomez2008}\\
Source I  & (05 35 14.5121, -05 22 30.521) & (6.3 $\pm$ 1.1, -4.2 $\pm$ 1.1) & (05 35 14.5163, -05 22 30.563) & \citealt{goddi2011}\\
BN Object  & (05 35 14.1131, -05 22 22.793) & (-3.5 $\pm$ 1.1, 10.4 $\pm$ 1.1) & (05 35 14.1253, -05 22 22.689) & \citealt{gomez2005}\\

\enddata
   \tablenotetext{a}{The average astrometric error of these proper-motion corrected positions is $\sim \pm$ 17 mas}

\end{deluxetable}

\subsection{Color Image Construction}

Color-temperature images were constructed using the \textit{L}$^{\prime}$ and \textit{Ms} images. The $L^{\prime}$ image was convolved with the $Ms$ PSF, and the $Ms$ image was convolved with the $L^{\prime}$ PSF; both PSFs were previously extracted by \textit{StarFinder}, and the resulting effective resolution in both maps was $\sim$ 135 mas. Images were then flux-calibrated using flux densities and aperture widths from \citet{gezari1998} ($Ms$, 360 Jy in a 3.9" $\times$ 3.9" aperture around BN) and \citet{dougados1993} ($L^{\prime}$, 1.54 Jy in a 0.75" $\times$ 0.75" aperture around Source n). The ratio of the PSF-matched $L^{\prime}$ and $Ms$ images was used to create a color-temperature map by assuming each pixel's flux ratio value was described by an independent blackbody. The uncertainties in the absolute photometry of the two images lead to an uncertainty of $\sim$ 31\% in the ratio map, and consequently in the color-temperature map, since the translation is roughly linear. This is a systematic uncertainty, and the relative differences in color-temperature between the pixels are much more precise than this. They are limited by the signal-to-noise ratio and PSF uncertainties. Neither extinction nor wavelength-dependent emissivity is included in the temperature determinations, so the derived color temperatures should not be interpreted as physical temperatures.

\section{Results}
\subsection{Detection of a Near-Infrared Feature Associated with Radio Source I}

The nature of Source I is of great interest because it is possibly the most powerful energy source in the closest region of massive star formation (see Section 1). In order to fully characterize Source I, it is necessary to elucidate the structure and dynamics of its disk.

We present the first detection of an infrared counterpart to the disk seen in the radio continuum associated with Source I in the bottom panels of Figure 3. We see a sharp intensity edge close to, and parallel with, the almost edge-on disk that is seen in the radio. This feature is most prominently seen in the color-temperature image made with the $L^{\prime}$ and $Ms$ bands (see section 3.3). The color-temperature of the surface of the counterpart visible in our NIRC2 data is lower than that of the ambient medium (see section 5.2 for discussion). The counterpart is not evident at $K^{\prime}$ due to high extinction (as high as $A_{v} = $ 58 magnitudes in the nearby IRc 2 region; \citealt{gezari1998}).  The emission in the $L^{\prime}$ and $Ms$ intensity maps that we interpret as arising from the surface of the disk implied by the radio data, while clearly present, is somewhat confused with source IRc 2A (see below and section 5.2). The semi-major axis of the disk counterpart is situated northwest to southeast, in agreement with the orientation of the circumstellar disk associated with Source I inferred by \citet{matthews2010}. There is also a hint of infrared emission from the outflow that is nearly perpendicular to the disk and that has been inferred from thermal SiO emission (see section 5.1). 

\subsection{Infrared Source IRc 2A}

The region containing the infrared counterpart to the Source I disk is a complex, extended source called IRc 2, in which several thermal infrared peaks have been identified (e.g., \citealt{dougados1993}). Most of the infrared radiation from IRc 2 at 2.2, 3.7, and 4.8 $\mu$m is extended in nature. All of the components of IRc 2 are resolved in our data except IRc 2C, which emits only about 10\% of the total radiation from IRc 2. The nearest component to Source I is IRc 2A, which we resolve into two separate components, IRc 2A-NE and IRc 2A-SW (see section 5). We speculate that these dust blobs are interlopers that lie within, or at the boundary of, the cone of illumination above the disk of Source I. Indeed, IRc 2A does not appear as a separate feature in the color-temperature map, suggesting that the dust within it is subject to the same radiation field as the dust in the contiguously projected flared disk of Source I. We are currently investigating the relative proper motions of IRc 2A and the Source I disk, and will present that in a separate publication.

\begin{figure}[!h]
\includegraphics[width=18cm]{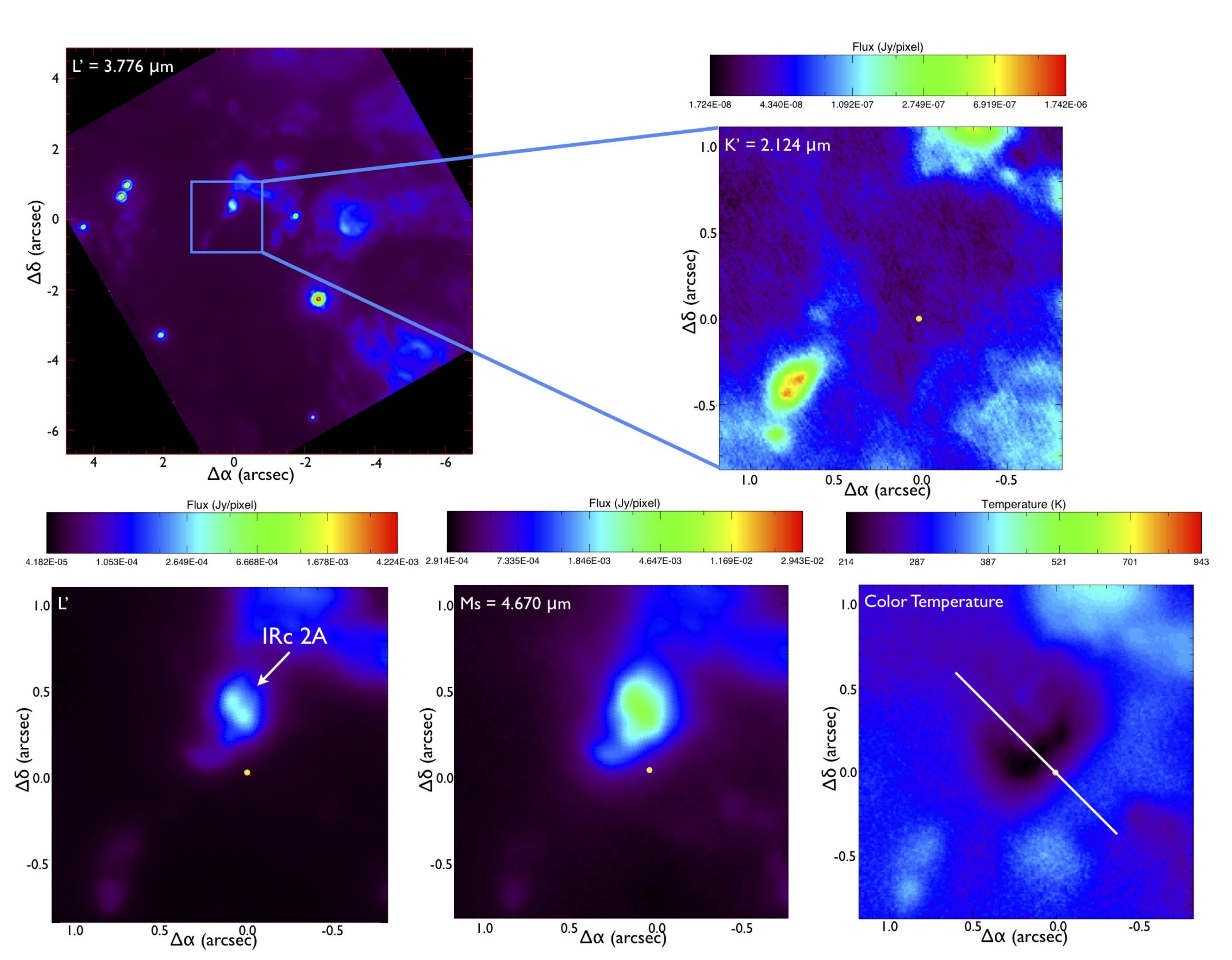}
\caption{NIRC2 narrow-field images in all three bands. The upper-left image shows our entire field of view at $L^{\prime}$ (see Figure 1); zooming into the boxed region, we see that this region is devoid of any emission at $K^{\prime}$ (upper right frame). However, in $L^{\prime}$ and $Ms$, we are able to see a peculiar morphology having a sharp lower edge. We interpret this feature and its sharp edge as the emission from the disk surface occulted by the opaque foreground portions of the disk. In the intensity maps, the disk appears to be confused by the infrared source IRc 2A, which is potentially an illuminated dust blob. All yellow dots denote the position of Radio Source I. The white line in the color-temperature map in the lower right frame corresponds to the intensity cut shown in Figure 5. The cut is oriented 45$^{\circ}$ east of north. The color-temperature map in this case corresponds to the blackbody-derived temperature between the $L^{\prime}$ and $Ms$ images.}
\end{figure}

\section{Discussion}
\subsection{Relationship of the Infrared Disk to the Elongated Radio Extension of Source I}

We interpret the sharp infrared edge as being the self-occulted, illuminated surface of the disk associated with Source I. This flared edge-on disk completely obscures the central protostar and inner disk, but the disk surface is illuminated in the flared outer regions and scatters starlight and emits thermally at the observed wavelengths (see Figure 6). The infrared counterpart to Source I manifest in the $L^{\prime}$ and $Ms$ images strongly suggests that we are seeing some combination of scattered or thermal light emanating from the disk.  

The outflow originating from Radio Source I has previously been indicated by the distribution of SiO maser emission \citep{matthews2010}, and the disk has been seen in 7-mm radio continuum (\citealt{goddi2011}; \citealt{reid2007}). Our detection of the upper surface of the inferred disk in both our color-temperature map and $L^{\prime}$ and $Ms$ intensity maps corresponds well with these previous radio studies.

Source I lies southwest of the infrared counterpart where the sharp edge in the infrared color-temperature map is at a position angle of $\sim 140^{\circ}$ and appears to be well-aligned with the projected radio disk major axis identified by \citet{reid2007} and \citet{goddi2011}. In order to compare the observations with radio observations of the disk, we over-plot the maser spots from \citet{matthews2010} (which arise from SiO $v = 1$ and $v = 2$, $J =$ 1 -- 0 emission) onto the $L^{\prime}$/$Ms$ color-temperature map in Figure 4. We propose that the offset of source I from the near-infrared edge seen in the NIRC2 data is caused by the obscuration of the region around the central star by the thick disk. The SiO maser emission spots are consistent with being at the inward (obscured) extension of the conical or flared surface of the disk that we see further out in the infrared (see middle panel of Figure 4). Furthermore, the darkest (coldest) region in the color-temperature map of Figure 4 lies slightly closer to the position of source I than the $L^{\prime}$ scattered light disk. This is a typical effect ascribable to the wavelength-dependent opacity of the circumstellar material lying in front of the disk itself \citep{watson2004, duchene2010, mccabe2011}, which suggests that the opacity of the dust in the circumstellar material between $L^{\prime}$ and $Ms$ is chromatic (i.e., non-grey). The models we present in section 5.2 fulfill this criterion. The right panel of Figure 4 shows the radio extension of Source I superposed on the color-temperature map of the visible portion of the disk. The contours represent 7-mm continuum intensity from the Very Large Array \citep{goddi2011} after subtraction of a Gaussian point source at the location of Source I. The continuum is well-aligned with the inferred disk edge depicted in the color map; the radio edge of the contours follows the darker-color region of our disk as seen in Figure 4. These 7-mm data presumably trace the dense mid-plane, whereas our near-infrared imaging traces the emission arising from the disk surface.

The offset of the edge of the disk to the position of Source I is reasonable, although higher than average, for an edge-on disk. The half-height of the disk is 67 $\pm$ 12 AU, estimated by measuring the distance between the position of source I and the lowest color temperature in the disk upper surface and assuming a nearly edge-on geometry. While there are no other known massive protostars with directly observed circumstellar disks, we can compare the Source I disk to those in lower mass systems. HH 30 has a $K$-band separation of 70 AU between the two reflection nebula \citep{cotera2000}; at $L^{\prime}$, HV Tau C has a full separation of only 33 AU \citep{duchene2010}, while HK Tau B has an even smaller separation of 23 AU \citep{mccabe2011}. An example of an intermediate-mass Herbig star with a known disk is PDS 144, but the distance to it is still highly uncertain so that no estimate is available for the physical separation of the two lobes of the nebula \citep{perrin2006}. 

The concave shape we observe presumably arises from the inner cavity wall being carved by the outflow. A hint of the outflow is present in the color-temperature map (Figure 4), where the emission appears cooler along the outflow axis relative to its immediate surroundings; presumably the counter jet is either not present or is obscured. A plot of surface brightness as a function of offset from Source I at $L^{\prime}$ along the outflow axis at an angle of $45^{\circ}$ east of north is shown in Figure 5. The absence of emission from the southwestern side of the disk, where the counterjet observed in thermal SiO emission is present, can be attributed to strong obscuration at wavelengths shorter than 5 $\mu$m.

\begin{figure}[!h]
\hskip-.8cm
\includegraphics[width=18cm]{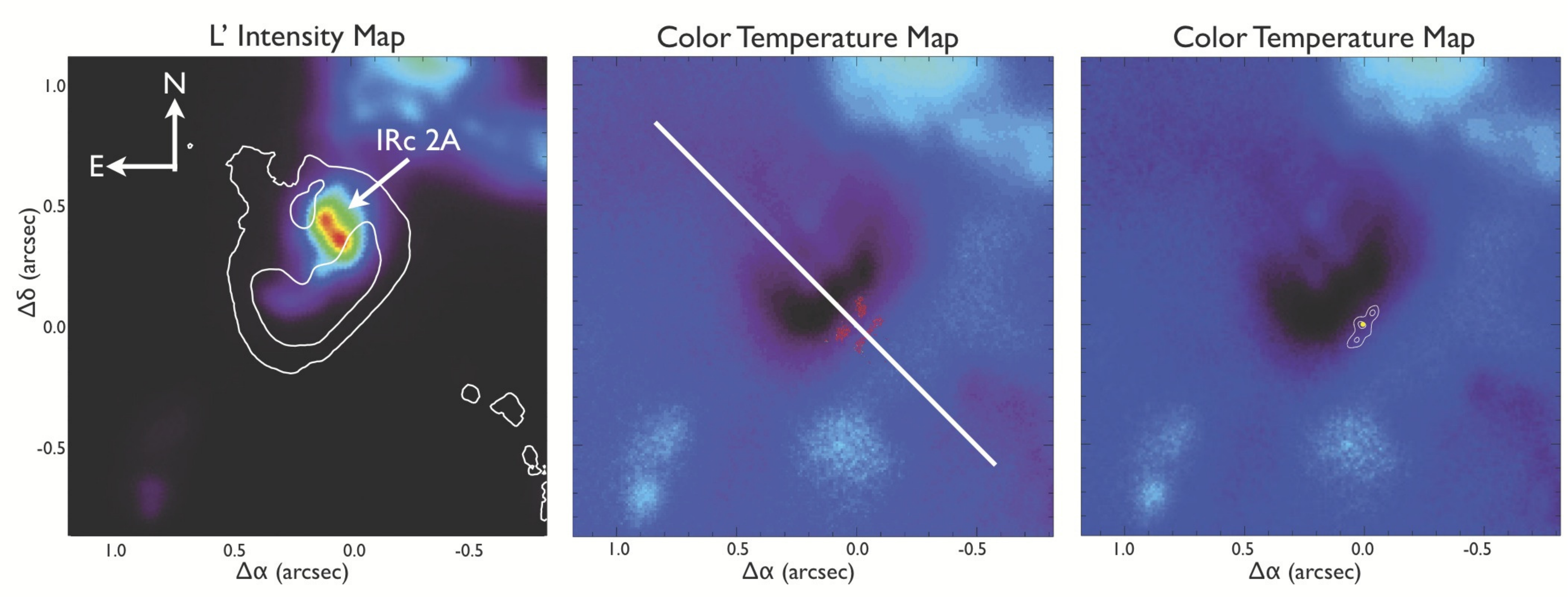}
\caption{Comparison of the infrared emission to various radio and sub-mm features associated with Source I's disk and outflow. \textit{Left}: $L^{\prime}$ intensity map with superimposed $L^{\prime}$/$Ms$ color-temperature map contours. \textit{Middle}: \citet{matthews2010} SiO maser emission spots overlaid on the color-temperature map. The alignment of the morphology of the infrared feature associated with Source I and the emission spots is clearly seen. The white line shows the orientation of the collimated thermal SiO outflow from \citet{plambeck2009}. There is also a relatively cool linear feature (compared to the disk) that is colinear with the star. It is not quite aligned with the SiO outflow, although the SiO outflow is not as well collimated. The disk itself is oriented at a PA of $\sim$ 125 $^{\circ}$ North of East. \textit{Right}: \citet{goddi2011} 7-mm continuum contours from the VLA superimposed on our color-temperature map. The flat surface of the proposed disk is parallel to the radio extension of Source I, as expected.}
\end{figure}

\begin{figure}[!h]
\begin{center}
\includegraphics[width=10cm]{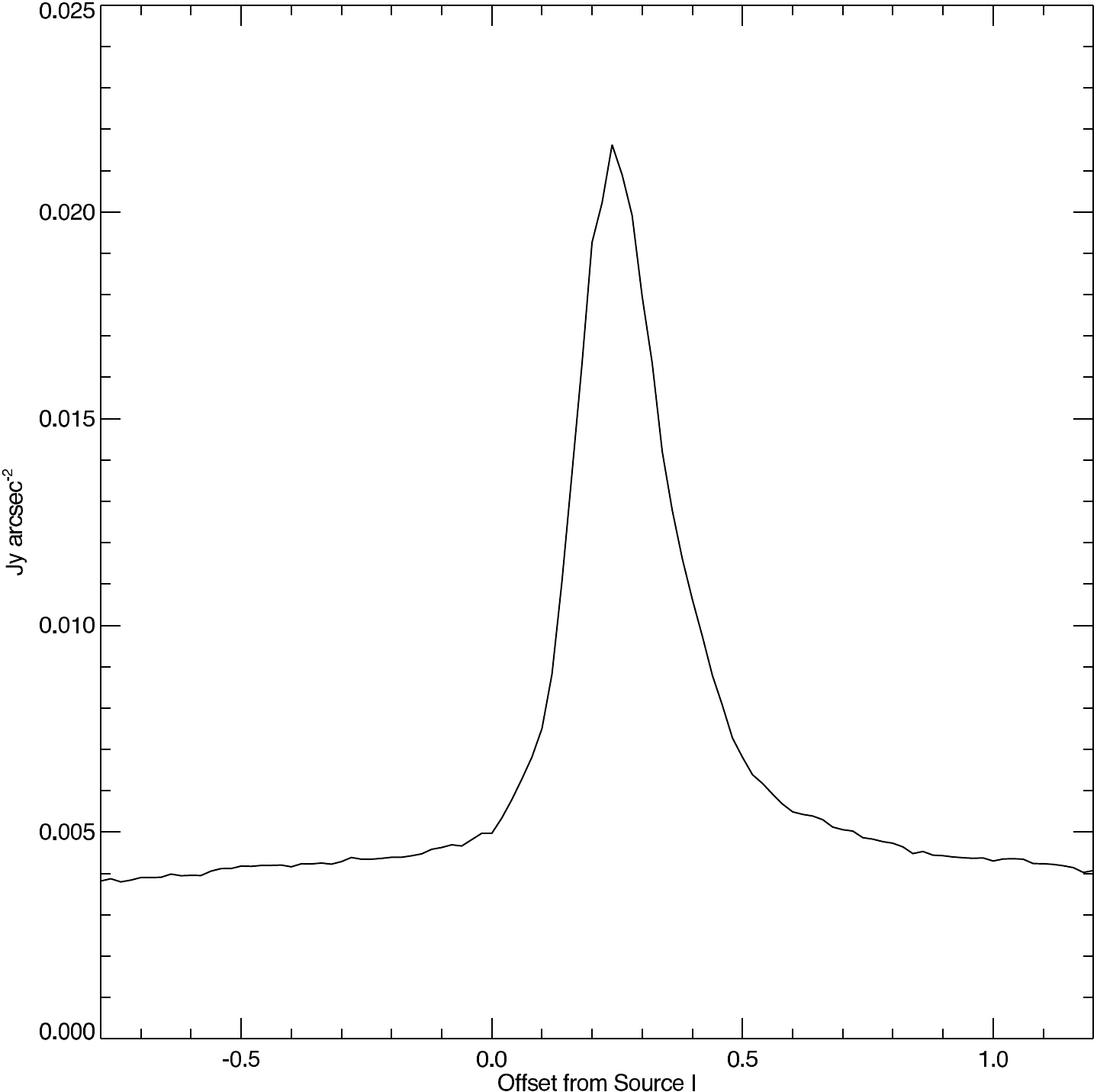}
\caption{Surface brightness as a function of offset from Source I at $L^{\prime}$. Note that no emission from a possible counter-jet is detected at negative offsets, but the sharp cutoff from what we infer to be the edge of the disk is evident. The position of the cut is shown in the bottom right-hand panel of Figure 3 and is at an angle of 45$^{\circ}$ east of north.}
\end{center}
\end{figure}

\subsection{Disk Models of the Data}
In order to explore the physical properties that govern the puzzling nature of Radio Source I, we have compared the morphology and colors of our infrared source with three-dimensional, Monte-Carlo radiative transfer and ray-tracing disk models using the MCFOST code \citep{pinte2006}. Since the SED of Source I remains unknown, as the source is enshrouded at all wavelengths blueward of radio wavelengths, the MCFOST modeling package characterizes both the dust population and gas disk properties exclusively with our near-infrared data constraints. 

\begin{figure}[!h]
\hskip -.8cm
\includegraphics[width=17cm]{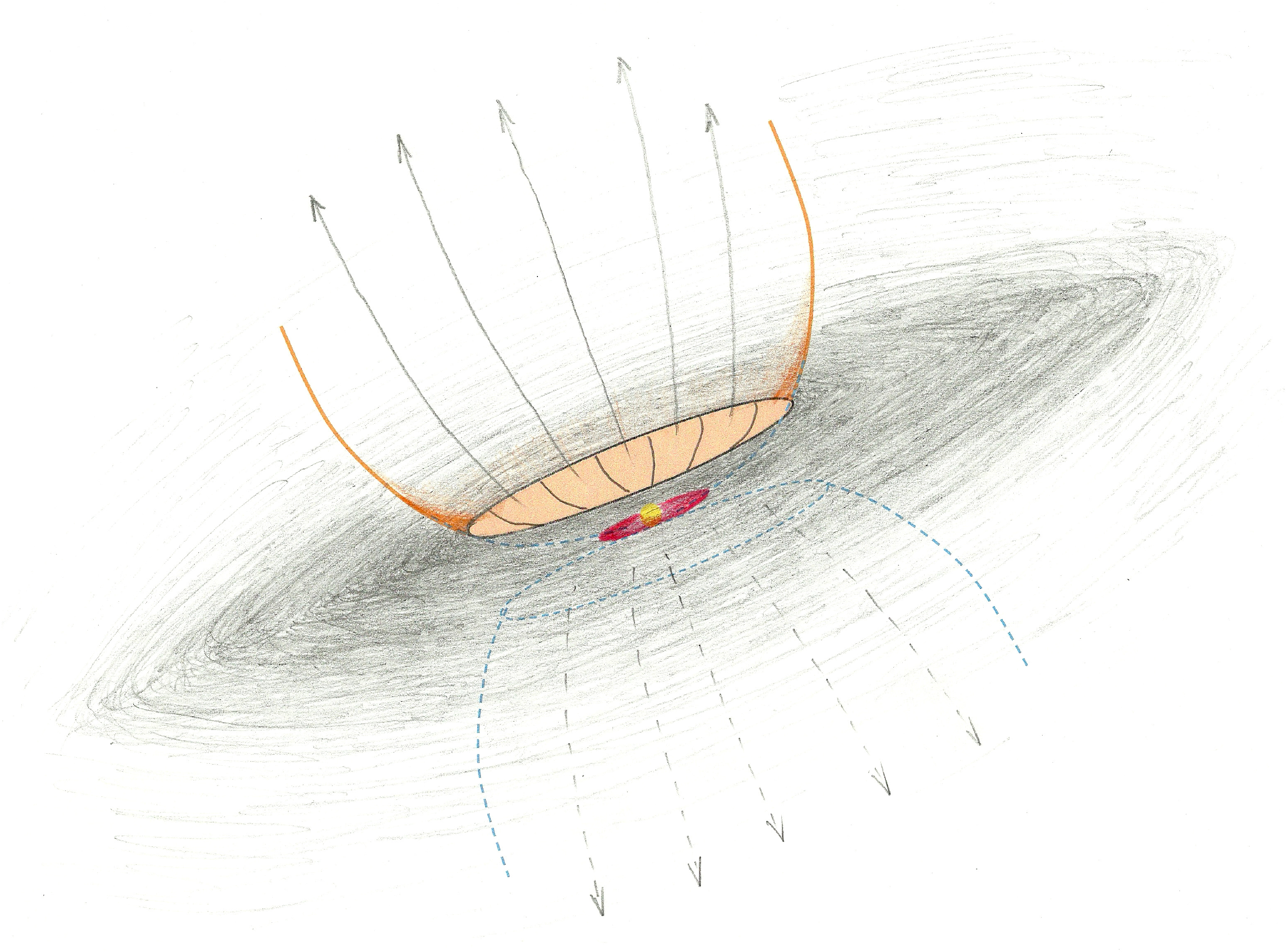}
\caption{A physical schematic of the disk around Source I. The star is denoted by a yellow dot, and the red color describes where the SiO maser emission occurs (e.g., \citealt{matthews2010}). The grey shaded area is the envelope and disk system accreting onto the star, while the  orange shows the bipolar outflow carving the cavity from the envelope. The light-orange color is the far side of the bubble wall illuminated by the star. Direct near-infrared emission from Source I is obscured by the high column density of the dust in the inferred disk.}
\end{figure} 

The models we used were divided into four sections that include:
\begin{itemize}
\renewcommand{\labelitemi}{$\bullet$}
\item A flared disk between 10 and 50 AU ($H/r \gtrsim 0.1$)
\item A spherically symmetric envelope from 10 to 400 AU
\item A geometrically flat and thick disk between 50 and 400 AU ($H = $ constant)
\item A paraboloidal cavity at the center of the spherical envelope, aligned with the symmetry axis of the disk ($H_{cav} = 20 (r/50)^{2}$ AU)
\item A cool central star ($T \sim$ 4,000 K and log($g$) = 3.5) with a large radius ($R \sim$ 300 \Rsun)
\end{itemize}
A schematic of the basic morphology is shown in Figure 6. The models utilized a Kurucz photospheric template consisting of a cool star. The near-infrared reflected spectrum of source I obtained at the VLT suggests that source I is a cool, bloated, massive protostar, which is why we used a cool photospheric temperature (\citealt{testi2010}; no spectrum was taken directly on Source I, only at the brightest locations of reflection nebulosity located at some distance from Source I).  The radius of the star was set to obtain a luminosity of $L \sim 21,000$ $\Lsun$ \citep{reid2007}, which was necessary to obtain a bolometric magnitude for the source appropriate for the massive protostar. There are no observational constraints on the luminosity due to the very high levels of extinction, so we have chosen characteristics appropriate for a massive protostar \citep{hosokawa2010}. However, it appears that most of the luminosity of the observed infrared source is not coming from the central high-mass star but from the base of the accretion flow and/or through re-radiation in a small cocoon around it. The outer radius was chosen to give about the right size to the overall structure, and the inner radius was set by the presumed $\sim$ 1,500 K sublimation temperature.

The placement of the SiO masers (see previous section) indicates that the disk must be flared out to at least 40 AU. At the surface of the outermost portion of the flared disk ($r$ = 50 AU), where the bulk of the emission that we observe arises, the model dust temperature is $\sim$ 600 K.

The symmetric, surrounding envelope was incorporated into the disk model in recognition of the fact that the source is likely embedded in a dense molecular cloud, and that the surrounding medium feeds material onto the disk. It was also necessary to generate the cavity walls well above the disk midplane and surface. Without the envelope, the near-infrared emission would be limited to the upper surface of the disk. The disk is flared; we see the outer edge, which is flat since all points where scattering occurs are at the outer radius of the disk, hence at a fixed $r$ and fixed $H(r)$. The flared disk is contiguous with the walls of the cylindrical cavity. The protoplanetary disk surrounding IRAS 04302+2247 also represents the early stages of stellar evolution and has been successfully modeled with an envelope surrounding it \citep{padgett1999, wolf2003}. CB 26 near the Taurus-Auriga star-forming region has been modeled with both a rotating, infalling envelope and a spherically symmetric density distribution. Figure 14 in \citet{sauter2009} compares the differences in these models. While both of these examples are low-mass analogues to Source I, \citet{perrin2006} also invokes a similar geometry to describe the environment around PDS 144N, an intermediate-mass system. Furthermore, introducing the cavity is necessary since we see ``walls" at the disk periphery in the color-temperature maps, and it is consistent with the outflow arising from Source I. The cavity is also consistent with previous simulations by \citet{cunningham2005}, as it produces a wide-angle outflow. The flat and thick disk between 50 and 400 AU acts as an opaque screen that attenuates the light from the southern-far-side nebula. This disk is a proxy for a dense midplane in an envelope that is presumably rotationally-supported.

\begin{figure}[!h]
\includegraphics[width=18cm]{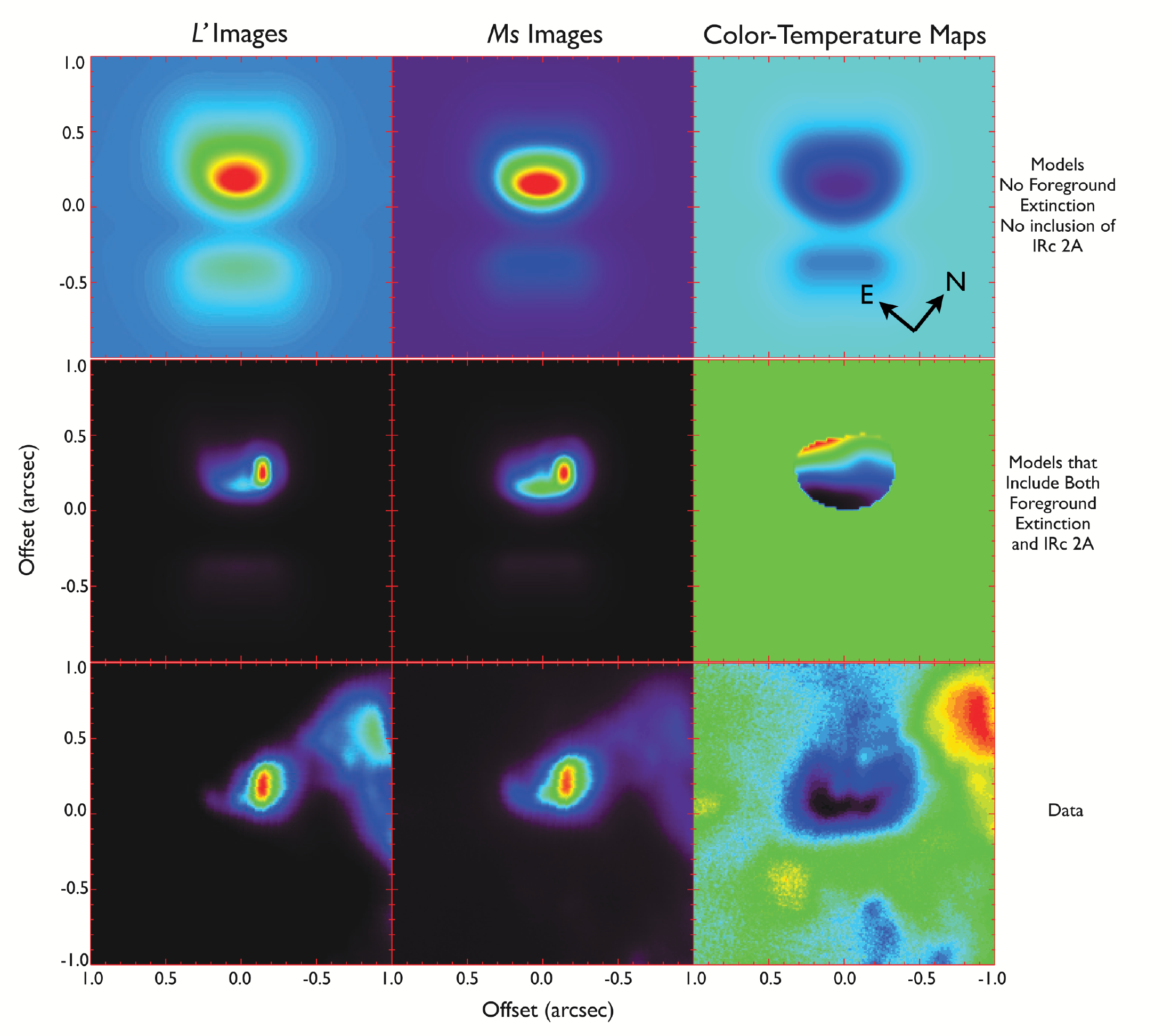}
\caption{MCFOST models displayed with our NIRC2 data for comparison. We convolved the $L^{\prime}$ and $Ms$ model images with their respective point spread functions extracted from our NIRC2 data, as generated by \textit{StarFinder} (see section 3.2). The ratio images were constructed as described in section 3.3. Each image shows a $2"\times 2"$ region. The top row shows the model images without the inclusion of foreground extinction or IRc 2A, while the middle row shows the same models including both foreground extinction and IRc 2A. We assumed IRc 2A consists of two separate point sources whose flux and position we have adjusted to roughly match observations. The bottom row shows our data from NIRC2.}
\end{figure}

Figure 7 depicts our disk models by presenting the intensities and the intensity ratio of the $L^{\prime}$ and $Ms$ data compared to those of the $L^{\prime}$ and $Ms$ model images with $i = 80^{\circ}$. We are able to reconstruct the basic morphology of the inferred disk with this model that contains both an envelope and disk component, namely a flat structure, and an obscured counternebula. The images in Figure 7 have been convolved with the respective PSFs extracted from \textit{StarFinder} (see section 3.2). The top row of the model images do not include any foreground extinction outside of the circumstellar envelope. A comparison with the data in Figure 7 shows how the region around our inferred disk surface has a different flux ratio than those of our models, which could arise from the lack of inclusion of the foreground extinction. If present, this will further redden the appearance of Source I and can account for the systematic offset in color between our observations and models. Studies from Spitzer \citep{zasowski2009, indebetouw2005} show that $A_{L^{\prime}}$/$A_{K}$ - $A_{Ms}$/$A_{K}$ $\sim$ 0.1. If $A_{v} \sim 60$ magnitudes of extinction \citep{greenhill2004a}, then $A_{K} \sim 6$ magnitudes, implying that there will be an extra 0.6 magnitudes in the observed $L^{\prime}$ - $Ms$ colors compared to our models. The middle row of Figure 7 includes this 0.6 magnitudes of foreground extinction, and includes the source IRc 2A. A much better fit to the data is obtained. We still have too much scattering well above the disk, which is a consequence of using a spherically symmetric envelope instead of a rotationally-supported envelope. A rotationally-supported envelope may produce a model with a more consistent gradient than what we depict with our spherically symmetric envelope.

We are seeing the disk in a combination of thermal and scattered light, along with foreground absorption from part of the envelope. The emission is mostly thermal, as scattered light from the central star only contributes $\sim$ 0.5\% of the total emission at $L^{\prime}$ and is relatively cool in the first place. The thermal emission favors longer wavelengths since the emitting dust is not very hot. Regions with colder color-temperatures (dark-colored in Figure 7) arise due to the wavelength-dependence of the dust opacity. At longer wavelengths like $Ms$, the optical depth through the envelope is lower. That is, the emitting region is embedded in the envelope and the $L^{\prime}$ photons are less likely to reach us than $Ms$ photons. Previous studies of grains in protoplanetary disks have shown evidence for grains significantly larger than those found in the interstellar medium, and several other systems show evidence for grains much larger than one micron (e.g., \citealt{shuping2006}; \citealt{guilloteau2008};  \citealt{lommen2010}; \citealt{melis2011}; \citealt{mccabe2011}). These large grains scatter efficiently in the near-infrared, so the photons thermally emitted in the innermost regions of the disk can thus be scattered off the outer edge of the disk without a strong blueing effect. In fact, our models have a maximum grain size of only one micron, yet our colors are still relatively red; our assumed dust population has an albedo $>$ 0.5 out to $\sim$ 4.8 microns. The source appears red because even grains of only 1 micron are quite effective scatterers out to $\sim$ 5 microns. Additionally, measuring near-infrared colors that are redder than those of the underlying star in the edge-on disk is further reinforced by the high stellar luminosity that heats the dust substantially, similarly to PDS 144N \citep{perrin2006}, and the presence of a surrounding, quasi-optically-thick envelope. The optically thick envelope produces reddening, as any photons scattered out of the light beam must pass through the envelope and cylindrical cavity.

\subsection{Source I as a merger product?}

The presumably massive star associated with Radio Source I has been considered to potentially be the product of a stellar merger \citep{bally2005} related to the violent dynamical interaction some 500 years ago of a non-hierarchal multiple system of massive protostars. According to this hypothesis, this interaction not only gave rise to the cataclysmic event of the NW-SE OMC 1 explosive outflow (``the fingers of Orion," \citealt{allen1993}), but also to the ejection and proper motion of Source I and Source n to the south and BN to the north \citep{gomez2005, zapata2009}. In that scenario, as \citet{bally2011} describe, Source I after the interaction would either be a close binary system of $\sim$10 solar mass objects with an orbital separation of the order of 1 AU or a bloated 20 $\Msun$ object resulting from the merger of two close binary components (\citealt{testi2010}; Portegies-Zwart, personal communication). The bloated star would relax to a smaller size and hotter temperature on a thermal Kelvin-Helmholtz timescale (estimated to be $\sim$10$^{4}$ years), but would still appear bloated now, as the suggested interaction would have taken place only 500 years ago.

The proposed interaction would likely have eliminated any pre-existing disk around Source I, so this model would require that the disk be reformed. A merger of the binary could provide the resulting massive star with substantial angular momentum, so an extruded, hot inner disk around the star might be expected. The \textit{outer} disk and dense envelope in this scenario would be the result of Bondi-Hoyle accretion onto the merger product (or tight binary) from a medium having a significant density or velocity gradient orthogonal to the stellar motion. This new gas is presumably captured from the dense surrounding medium (OMC1 hot core, density $\sim$10$^{6}$ cm$^{-3}$) into which it is hypothesized to be moving at a relative speed of about 10 km s$^{-1}$. Moreover, the model posits that Source I is moving close to the plane of the sky and that the angular momentum vector of the newly accreted disk would lie in a plane orthogonal to the stellar velocity vector, so that the apparent major axis of such a disk as projected onto the plane of the sky will be aligned with the proper motion vector. Thus, while the model does not dictate that this new disk of Source I also be oriented edge-on, the predicted disk orientation is completely consistent with the observed 7-mm VLA disk structure, and the collimated outflow associated with the thermal and maser SiO emission.

Whether this speculative dynamical scenario is applicable and can be reconciled with our Keck/NIRC2 observations, which point to a more regular rather than a presumably chaotic disk/envelope structure, is an open question. Time-dependent relaxation phenomena may be observable in the future to support the dynamical scenario.

\section{Conclusions}
We have detected a near-infrared counterpart to Radio Source I in the BN/KL region of Orion in our NIRC2 $L^{\prime}$, $Ms$, and $L^{\prime}/Ms$ color-temperature maps. We interpret this counterpart as the protostellar disk around Radio Source I. In the intensity maps, it is confused by the infrared source IRc 2A, the nearby near-infrared peak. The morphology of the extended disk emission is also prominently traced in the color-temperature map. The orientation and the nearly edge-on aspect of the infrared emission from the disk are consistent with the 7-mm continuum observations of \citet{goddi2011}. Furthermore, the morphology and flared behavior of the disk are very consistent with the geometry suggested by the SiO maser emission spots from \citet{matthews2010}. 

We have modeled the near-infrared counterpart to Source I with a full dust radiative transfer model that includes an envelope. Our models indicate that we are seeing the disk in both thermal emission from particles heated by Source I, along with scattered light from Source I. The disk appears red in our ratio images because we are looking at part of the emission through the foreground screen that arises from the wall of the cylindrical cavity. Light scattering through this screen causes differential extinction, as discussed in section 5.2. Our Keck observations coupled with our MCFOST models support the hypothesis that the morphological shape we see above Source I in our color-temperature map is a consequence of disk emission.

Further progress in understanding the disk of Source I can be made when direct measurements of line emission become possible. If the disk is ionized, as \citet{goddi2011} and \citet{reid2007} have argued, then interferometric measurements of radio recombination lines might be valuable, and if it remains predominantly molecular, then ALMA should eventually reveal its constituents and dynamics.

\vspace{10mm}

Support for this work was provided by a Universities Space Research Association (USRA) grant to UCLA (grant USRA850005). A. S. acknowledges support from the German Science Foundation Emmy Noether program under grant DFG STO 496-3. J. R. L. acknowledges support from the NSF Astronomy and Astrophysics Postdoctoral Fellow program (AST -1102791). 

Data presented herein were taken at the W. M. Keck Observatory. The W. M. Keck Observatory is operated as a scientific partnership among the California Institute of Technology, the University of California, and the National Aeronautics and Space Administration. The Observatory was made possible by the generous financial support of the W. M. Keck Foundation. We wish to acknowledge the significant cultural role that the summit of Mauna Kea has always had for the indigenous Hawaiian community. We are most fortunate to have the opportunity to conduct observations from this mountain. 

B. N. S. would like to thank Shane Frewen and Sylvana Yelda for their time and assistance. The authors wish to acknowledge the invaluable comments from the anonymous referee. 

\textit{Facilities}: Keck: II (NIRC2)

\end{document}